\documentclass[12pt]{iopart}

\usepackage{graphicx}
\begin{document}

\title[]{Mini Black Holes in the first year of the LHC \\
{\small Discovery Through Di-Jet Suppression, Mono-Jet Emission and ionising tracks in ALICE}}

\author{ H.
St\"ocker$^{1,2}$ }
{\normalsize
\hspace*{-8pt}$^1$ FIAS- Frankfurt Institute for Advanced Studies\\
D--60438 Frankfurt am Main, Germany\\
\hspace*{-8pt}$^2$ Institut f\"{u}r Theoretische Physik, Johann Wolfgang Goethe -
Universit\"{a}t\\
D--60438 Frankfurt am Main, Germany\\[0.2ex]
}

\ead{stoecker@fias.uni-frankfurt.de}
\begin{abstract}
The experimental signatures of TeV-mass black 
hole (BH) formation in heavy ion collisions at the LHC is examined. 
We find that the black hole production results in a complete disappearance 
of all very high $p_T$ (\mbox{$> 500$} GeV) back-to-back correlated 
di-jets of total mass \mbox{$M > M_f \sim 1$}TeV. We show that the 
subsequent Hawking-decay produces multiple hard mono-jets and discuss their detection.
We study the possibility of cold black hole remnant (BHR) formation of mass $\sim M_f$
and the experimental distinguishability of scenarios with BHRs and those with complete black hole decay.
Due to the rather moderate luminosity in the first year of LHC running the least chance for the observation
of BHs or BHRs at this early stage will be by ionizing tracks in the ALICE TPC. 
Finally we point out that 
stable BHRs would be interesting candidates for energy production by conversion
of mass to Hawking radiation.
\end{abstract}

\maketitle

\section{Introduction}\label{intro}

Frankfurt-born astronomer Karl Schwarzschild discovered the first analytic solution of the General Theory of Relativity \cite{Schwarzschild}. 
He laid the ground for studies of some of the most fascinating 
and mysterious objects in the universe: 
the black holes. Recently, it was conjectured that black holes (BH) do also reach into the regime of particle physics: 
In the presence of additional
compactified large extra dimensions (LXDs), it seems possible to produce tiny 
black holes in colliders such as the Large Hadron 
Collider (LHC), at the European Center for Nuclear Research, CERN. 
This would allow for tests of Planck-scale physics and of the onset 
of quantum gravity - in the laboratory! Understanding black hole physics is a key to the phenomenology of these new effects beyond the 
Standard Model (SM).

During the last decade, several models \cite{Antoniadis:1990ew,add,rs1} using extra dimensions as an additional assumption to the 
quantum
field theories of the Standard Model (SM) have been proposed. The most intriguing feature of these models is that they provide a 
solution
to the so-called hierarchy problem by identifying the "observed" huge 
Planck-scale as a geometrical feature of the space-time,
while the true fundamental scale of gravity $M_f$ may be as low as 1 TeV. 
The setup of these effective models is partly motivated by String
Theory. The question whether our space-time has additional dimensions is well-founded on its own and worth the effort of 
examination. 

In our further discussion, we use the model proposed by Arkani-Hamed, Dimopoulos and Dvali
\cite{add}, proposing $d$ extra space-like dimensions without curvature, each of them compactified to a certain radius $R$. 
Here all SM 
particles are confined to our 3+1-dimensional brane, while gravitons are allowed to propagate freely in the (3+d)+1-dimensional bulk. 
The Planck mass $m_{Pl}$ and the fundamental mass $M_f$ are related by
\begin{eqnarray}
m_{Pl}^2 = M_f^{d+2} R^d \quad. \label{Master}
\end{eqnarray}

The radius $R$ of these extra dimensions can be estimated using Eq.(\ref{Master}). 
For $d$ equaling $2$ to $7$ and $M_f \sim$~TeV, $R$ extends from $2$~mm to $\sim 10$~fm.
Therefore, the inverse compactification radius $1/R$ lies in energy range 
eV to MeV, respectively. The case $d=1$ is excluded: It would result in an extra dimension about the size of the solar system. For 
recent 
updates on constraints on the parameters $d$ and $M_f$ see e.g. Ref.\ \cite{Cheung:2004ab}.

\section{Estimates of LXD-black hole formation cross sections at the LHC}

The most exciting signature of {\sc LXD}s is the possibility of black hole production in colliders 
\cite{Banks:1999gd,dim,ehm,Giddings3,Hofmann:2001pz,Hossenfelder:2001dn,Hossenfelder:2003jz,Hossenfelder:2003dy,Casadio:2001dc,Alexeyev:2002tg,BleicherNeu,Chamblin:2002ad,Casanova:2005id,own2,Casadio:2001wh,Hossenfelder:2004ze,Stojkovic:2004hp,Hossenfelder:2005bd,Hst06,Hst062,Alberghi:2006qr,Betz:2006ds} 
and 
in ultra high energetic cosmic ray events \cite{cosmicrayskk,cosmicraysbh}: At distances below the size of the extra
dimensions the Schwarzschild radius \cite{my} is given by
\begin{equation} \label{ssradD}
R_H^{d+1}=
\frac{2}{d+1}\left(\frac{1}{M_{f}}\right)^{d+1} \; \frac{M}{M_{f}}
\quad .
\end{equation}
This radius is much larger than the corresponding radius in 3+1 dimensions. Accordingly,
the impact parameter at which colliding particles form a black hole via the 
Hoop conjecture \cite{hoop} rises enormously in the extra-dimensional
setup.
The LXD-black hole production cross section can be approximated by the classical geometric 
cross section
\begin{eqnarray} \label{cross}
\sigma(M)\approx \pi R_H^2 \quad,
\end{eqnarray}
which only contains the fundamental scale as a coupling constant.

This classical cross section has been under debate
\cite{Voloshin:2001fe,Rychkov:2004sf,Jevicki:2002fq}: Semi-classical considerations 
yield form factors of order one 
\cite{Formfactors}, which take into account that only a fraction of the
initial energy can be captured behind the Schwarzschild-horizon. 
Angular momentum 
considerations change the results by a factor 
of two \cite{Solo}. Nevertheless, the naive classical result 
remains valid also in String Theory \cite{Polchi}.

\begin{figure}[htb]
\begin{minipage}[c]{6 cm}
\vspace*{-.1cm}
\begin{center}
\includegraphics[width=5.5truecm]{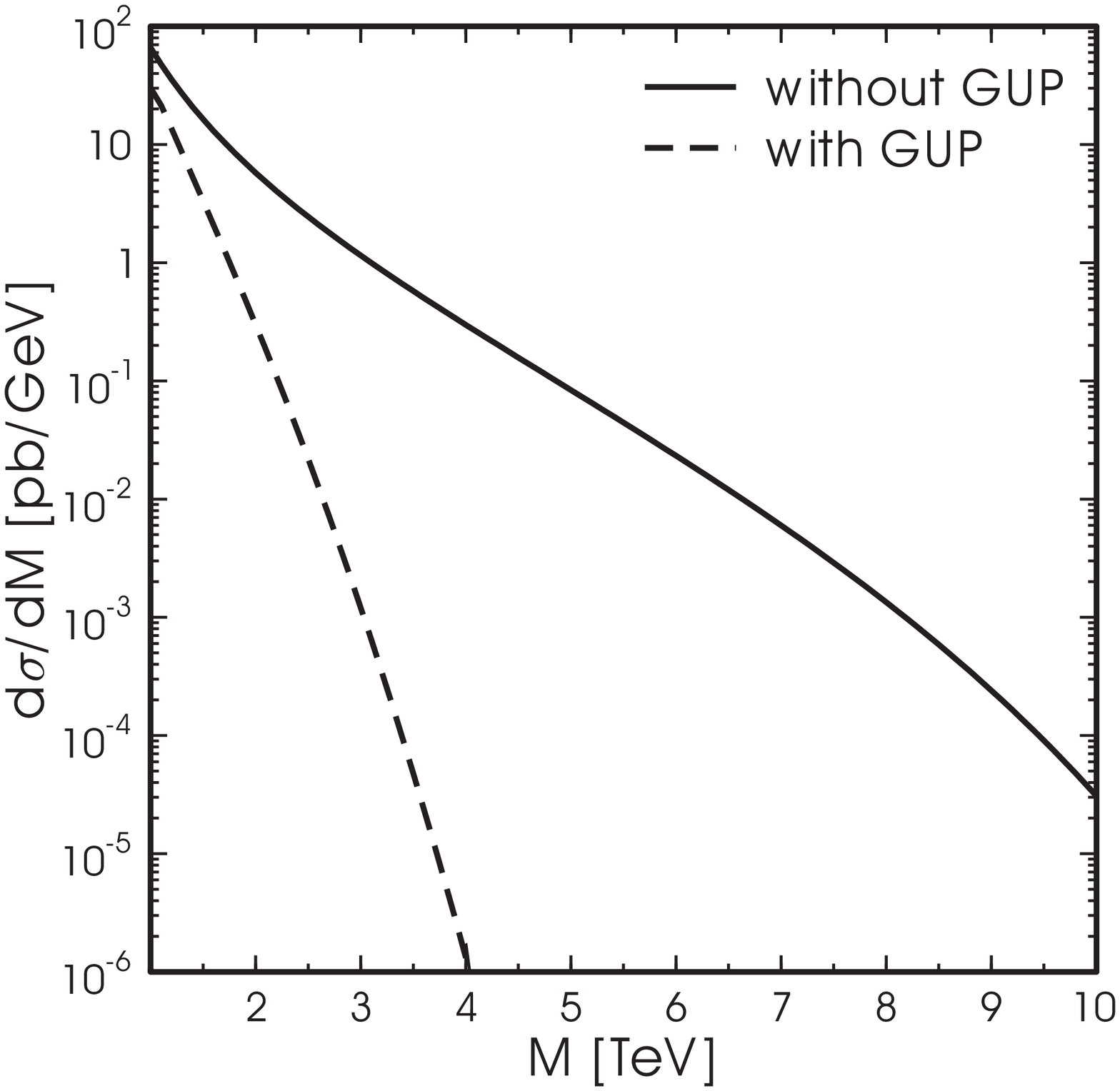}
\end{center}
\end{minipage}
\begin{minipage}[c]{6 cm}
\vspace*{-.1cm}
\begin{center}
\includegraphics[width=5.5truecm]{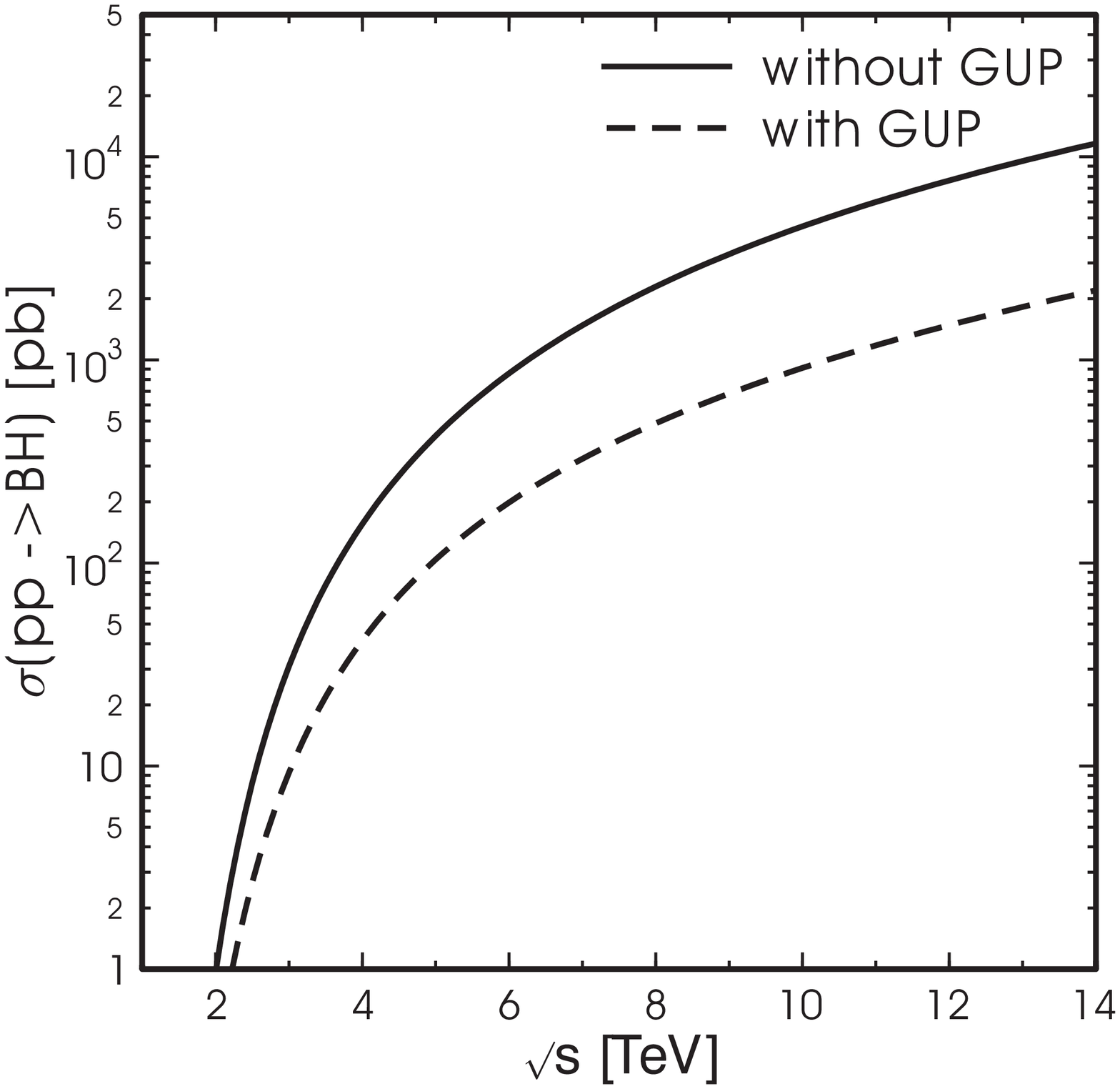}
\end{center}
\end{minipage}
\caption{The left plot shows the differential cross section for black hole
production in p-p collisions at $\sqrt{s}=14\rm{TeV}$ ({\sc LHC}) for $M_{f}=1$~TeV. 
The right plot shows the integrated cross section for 
BH production as a function of the 
collision energy $\sqrt{s}$. In both cases, the curves for various
$d$ differ only slightly from the above depicted ones.
The dashed curves show calculations including the minimal length
(via a Generalized Uncertainty Principle (GUP)) {\protect\cite{Hossenfelder:2004ze,Hossenfelder:2005bd}}.}
\protect{\label{dsdm}}
\end{figure}

Stronger modifications to the BH cross section are expected from recent calculations introducing a minimal length scale,
suggested by String Theory and Loop Quantum Gravity alike. Via the use of a model implementing a
Generalized Uncertainty Principle (GUP), one can show 
that a minimal length scale leads to a reduction of the density 
of states in momentum space at high energies. The squeezing of the
momentum states not only reduces the 
black hole cross section, but also Standard Model cross sections involving high momentum transfer
\cite{Hossenfelder:2004ze}, see Fig. \ref{dsdm}.
\vspace*{1ex}

Setting $M_f\sim 1~$TeV and $d=2 - 7$ one finds cross-sections of $\sigma \sim 400$~pb$ - 10$~nb. Using the geometrical 
cross section formula, it is now possible to compute the differential cross section ${\mathrm d}\sigma/{\mathrm d}M$ for p-p collisions 
with an invariant energy $\sqrt{s}$. This cross section is given by the
summation over all possible parton interactions and integration over the momentum fractions $x_i$, where  the kinematic relation 
$x_1 x_2 s=\hat{s}=M^2$ has to be fulfilled. This yields the expression
\begin{eqnarray} \label{partcross}
\frac{{\rm d}\sigma}{{\rm d}M}
&=&  \sum_{A, B} \int_{0}^{1} {\rm d} x_1 \frac{2 \sqrt{\hat{s}}}{x_1s}
f_A(x_1,\hat{s})
f_B (x_2,\hat{s})  \sigma(M,d).   \quad
\end{eqnarray}

A numerical evaluation \cite{Hossenfelder:2005bd} using the 
{\sc CTEQ4}-parton distributions $f_i(x,Q)$ results in the cross section 
displayed in Figure \ref{dsdm}. 

One can see that independent of the specific scenario, most of the black holes created have masses 
close to the production threshold. This is due to the fact that the parton distribution functions $f_i(x_i)$ are strongly peaked at 
small values of the momentum fractions $x_i$.

At the LHC up to $10^9$ black holes may be created per year with the 
estimated full LHC luminosity of $L=10^{34}{\rm cm}^{-2}{\rm s}^{-1}$ at $\sqrt{s}=14$~TeV: Depending on the specific scenario, about 
ten black holes per second could be created \cite{dim}.
In the first year of running this rate will be thousand fold lower.
LXD-black hole production would have dramatic consequences for future collider physics: Once the collision energy crosses the threshold 
for black hole production, no further information about the structure of matter at small scales can be extracted - 
this would be ''{\it{the end of short distance physics}}'' \cite{Giddings3}.


\section {Suppression of high mass correlated di-jet signals in heavy ion collisions}

The above findings led to a high number of publications on the topic of 
TeV-mass black holes at colliders
\cite{dim,ehm,Giddings3,Hofmann:2001pz,Hossenfelder:2001dn,Chamblin:2002ad,own2,Casadio:2001wh,Hst06,Hst062,Landsberg:2002sa,Marcus,Lonnblad:2005ah,Hossenfelder:2005ku,Humanic,PYTHIA,CHARYBDIS,BHex,Koch:2005ks}, 
for hadronic collisions as well as for heavy ion 
collisions \cite{own2,Hst06,Chamblin:2003wg}: 
At the same center of mass energy, the number of black holes in a heavy ion event 
compared to a hadronic event is increased by about
thousandfold due to the scaling with the number of binary collisions \cite{Chamblin:2003wg}.

The first, cleanest signal for LXD-black hole formation in Pb-Pb collisions 
is the complete suppression of high energy back-to-back-correlated 
di-jets with $M > M_f$: two very high energy partons
which usually define the di-jets in the Standard Model, each having an energy of 
$\sim$ one-half $M_f$ (i.e. $p_T \geq 500$ GeV), now end up inside the black 
hole \cite{own2,Casadio:2001wh,Hst06,Lonnblad:2005ah} instead of being observable 
in the detector.
Di-jets with $E_{di-jet} > M_f$ cannot be emitted.


\section{Hard, isotropic multiple mono-jet emission as a signal for hot LXD-black hole hawking-evaporation}

Once produced, the black holes may undergo an evaporation process
\cite{Hawk1}
whose thermal properties carry information about the parameters $M_{\rm f}$ and $d$. An
analysis of the evaporation
will therefore offer the possibility to extract knowledge about the
topology of space time and the underlying theory.

To understand the signature caused by black hole decay,
we have to examine the Hawking evaporation process in detail:
The evaporation rate ${\rm d}M/{\rm d}t$ can be
computed for an arbitrary number of dimensions using the thermodynamics of black holes. 
The Hawking-temperature ($T$) depends on the black hole radius
\begin{eqnarray} \label{tempD}
T=\frac{1+d}{4 \pi}\frac{1}{R_H} \quad,
\end{eqnarray}
which is given by Eq. (\ref{ssradD}).
The smaller the black hole, the larger its temperature.

Integrating the thermodynamic identity d$S/$d$M = 1/T$ over $M$ yields 
the entropy
\begin{eqnarray}
S(M)
&=& 2 \pi  \frac{d+1}{d+2} \left( M_f R_H \right)^{d+2}\quad.
\end{eqnarray}
With rising temperature,
the emission of a particle will have a non-negligible
influence on
the total energy of the black hole. This problem can appropriately be
addressed by
including the back-reaction of the emitted quanta as derived in Ref.\ \cite{Page,backreaction}. It is found that in the regime of
interest, when
$M$ is of order $M_{f}$, the number density
for a single particle micro state $n(\omega)$ is modified and now given by the change
of the black
hole's entropy:
\begin{equation} \label{nsingle}
n(\omega) =  \frac{\exp[S(M-\omega)]}{\exp[S(M)]}\quad .
\end{equation}
From this, using the evaporation rate we obtain
\begin{eqnarray} \label{mdoteq}
\frac{{\mathrm d}M}{{\mathrm d}t} = \frac{\Omega_{(3)}^2}{(2\pi)^{3}}
R_H^{2} 
\int_{0}^{M} 
\frac{\omega^3 \,\rm{d}\omega}{\rm{exp}\left[S(M-\omega)-S(M)\right]}\quad,
\end{eqnarray}
where $\Omega_{(3)}$ is the 3-dimensional unit sphere.

One observes that the evaporation process
of the black holes slows down
in its late stages \cite{Hossenfelder:2001dn,Hossenfelder:2003jz}
\footnote{In a 3-dimensional theory this enhanced lifetime can
also be obtained from a renormalization group 
approach \cite{Bonanno:2006eu}.}, 
and may even come to a complete stop, thus, stable black hole remnants may be 
formed 
\cite{Hossenfelder:2003jz,Hst06,Hst062,Bonanno:2006eu,Koch:2005ks}.


The above discussion allows for the following observations:
\begin{itemize}
\item Typical temperatures for LXD-black holes with $M_{BH} \gg M_f$, e.g. $5-10$~TeV,
are several hundred GeV. This high temperature results in a very short lifetime. The black
hole will decay close to the primary interaction region and thus its decay products can be observed
in collider detectors.
\item Most of the SM particles of the black body radiation are emitted with $\sim 100$~GeV average
energy, which leads to multiple high energy mono-jets with much higher multiplicity than in Standard Model
processes \cite{Hst06}.
\item The total number of emitted jets can be estimated to be of order $10$. Because of the thermal
characteristics of the decay, the pattern will be nearly isotropic, with  a high sphericity of the event.
\end{itemize}

Although the high mass BHs might give the cleanest signatures, one
has to keep in mind that 
in the first year of LHC running
one has to search for BHs or BHRs in the low mass region
(slightly above 1TeV) as the rather moderate luminosity will only 
allow for the production of a small number of those objects which have most likely
masses just above the production threshold.

Ideally, the energy distribution of the decay products allows for a
determination of the temperature (by fitting the energy spectrum to the
predicted shape) as well as of the total mass of the BH (by summing up all
energies).
This then will allow for a reconstruction of the fundamental scale $M_{\rm f}$ and the
number of extra dimensions.

Several experimental
groups have included LXD-black hole searches into their research programs for
physics beyond the Standard Model, in particular the ALICE, ATLAS and 
CMS collaborations at the LHC \cite{Humanic}. PYTHIA 6.2 \cite{PYTHIA} 
with the CHARYBDIS \cite{CHARYBDIS} event generator allows for a simulation of
black hole events and data reconstruction from the
decay products. Such analysis has been summarized in Refs. \cite{Humanic,Atlas,BHex}.
If only low mass BHRs are produced, however, these signals don«t exist. Therefore
one has to search for the stable ($M=1$TeV) ionizionizinging track of the BHR
in the ALICE TPC.

\section{Formation of stable black hole remnants and single track detection in the ALICE-TPC}

To obtain predictions for collider experiments, one has to produce numerical simulations incorporating black hole events. 
These simulations have been performed but have so far assumed mostly that the black holes decay completely into Standard 
Model particles. 
As already pointed out, however, there are equally strong indications that the black holes do NOT
evaporate completely, but rather leave a meta-stable black hole remnant
(BHR) \cite{Hossenfelder:2001dn,Hossenfelder:2003jz,Hossenfelder:2003dy,Hst06,Hst062,Bonanno:2006eu,Koch:2005ks}.

\begin{figure*}[htb]
\includegraphics[width=10.5truecm]{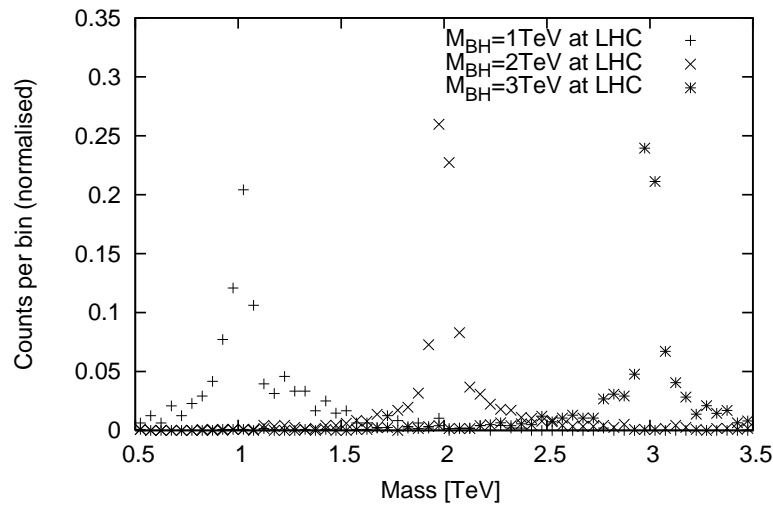}
\caption{
Reconstructed BHR masses in p-p reactions at $\sqrt{s}=14$ TeV from ALICE (TOF $56$ ps) resolution
for $M_{\rm BH}=1,2,3$ TeV\label{Bhm}\cite{Betz:2006ds,Humanic}.
}
\end{figure*}

If BHRs are formed, they can carry charge and may thus not only be reconstructed via decay products, but can rather 
directly be observed: 
Charged BHRs should appear in 
the ALICE detector at the LHC as a magnetically very
stiff charged (small curvature) track.
As shown in Fig. \ref{Bhm}, the mass of a charged BHR can be reconstructed
within the ALICE time of flight and spatial resolution \cite{Humanic}. 


\section{Black hole remnants as interesting candidates for energy production by conversion
of mass to Hawking radiation}

If stable BHRs really exist one could  not only study them with various experimental
setups but also use them
as catalyzers to capture and convert, in accordance with  
$E=mc^2$, high intensity beams of low 
energy baryons (p,n, nuclei), of mass $\sim 1$ AGeV, into photonic, leptonic and light
mesonic Hawking radiation, thus serving 
as a source of energy with 90\% efficiency (as only neutrinos and 
gravitons would escape the detector/reactor).
If BHRs (Stable Remnants) are made available by the LHC or the NLC and can 
be used to convert mass in energy, then the 
total 2050 yearly world energy consumption of roughly $10^{21}$ Joule can be 
covered by just $\sim 10$ tons of arbitrary material, 
converted to radiation by the Hawking process via $m = E/c^2 = 10^{21} 
\hbox{J}/(3\cdot 10^8 \hbox{m/s})^2 = 10^4$ kg 
\cite{HorstPatent}. 

\section{Conclusion}
The LHC will provide exciting discovery potential way beyond supersymmetric extensions of the Standard Model. 
Still one has to keep in mind that the LHC will run in the first year with rather
moderate luminosity. 
Hence we first must focus on the dominant part of the production cross section for BHs,
which is just slightly above the production threshold.
The most prominent signatures in this regime
are a complete suppression of back-to-back correlated di-jets,
the production of mono-jets with energies $< 1$TeV
and the possibility of the formation of stable black hole remnants. 
We have shown how signatures in the ALICE TPC chamber can be used to 
identify BHRs with masses of $\approx 1$TeV in the first year of LHC running,
even at rather moderate luminosities.

\section*{Acknowledgements}
This work has been supported by GSI, BMBF and the ALICE collaboration.
Special thanks to
Harry Appelsh\"auser, Peter Braun-Munzinger, 
Johanna Stachel, Sabine Hossenfelder, Ben Koch and Tom Humanic.

\end{document}